\documentclass[letterpaper]{article}

\usepackage[english]{babel}
\usepackage[utf8]{inputenc}
\usepackage[T1]{fontenc}
\usepackage[normalem]{ulem}

\usepackage[letterpaper,top=3cm,bottom=2cm,left=3cm,right=3cm,marginparwidth=1.75cm]{geometry}

\usepackage{amsmath}
\usepackage{graphicx}
\usepackage[colorinlistoftodos]{todonotes}
\usepackage[colorlinks=true, allcolors=blue]{hyperref}
\usepackage{siunitx}
\usepackage{threeparttable}
\usepackage{indentfirst}

\title{ENHANCED DIFFUSION AND CHEMOTAXIS OF ENZYMES}
\author{Mudong Feng and Michael K. Gilson}

\begin{document}
\maketitle
\begin{center}Department of Chemistry and Biochemistry, and Skaggs School of Pharmacy and Pharmaceutical Sciences, UC San Diego, La Jolla, CA 92093\end{center}

(When citing this paper, please use the following: Feng M, Gilson MK. 2019. Enhanced Diffusion and Chemotaxis of Enzyems. Annu. Rev. Biophys. 49: Submitted.)

\section*{Abstract}
Many enzymes appear to diffuse faster in the presence of substrate and to drift either up or down a concentration gradient of their substrate. Observations of these phenomena, termed enhanced enzyme diffusion (EED) and enzyme chemotaxis, respectively, lead to a novel view of enzymes as active matter. Enzyme chemotaxis and EED may be important in biology, and they could have practical applications in biotechnology and nanotechnology. They also are of considerable biophysical interest; indeed, their physical mechanisms are still quite uncertain. This review provides an analytic summary of experimental studies of these phenomena and of the mechanisms that have been proposed to explain them, and offers a perspective of future directions for the field.

\section*{Keywords}
Fokker-Planck equation, microfluidics, FCS, urease, catalase, aldolase

\tableofcontents

\section*{DEFINITIONS}
\begin{description}
    \item[Enhanced Enzyme Diffusion (EED)] An apparent speedup of translational enzyme diffusion. Usually reported as an increase in the diffusion coefficient.
     \item[Active Mechanism] A proposed mechanism of EED or chemotaxis is active if it requires the release of chemical energy by enzyme catalysis.
    \item[Enzyme Chemotaxis] Apparent drift of an enzyme up (attractive) or down (repulsive) a chemical gradient.\end{description}

\section{INTRODUCTION}\label{intro}
Studies over the last decade suggest that non-motor enzymes engage in active, translational motion. Two phenomena have been noted. One, termed enhanced enzyme diffusion (EED), is an increase in the enzyme’s diffusion constant induced, typically, by provision of substrate. The other, enzyme chemotaxis, is a tendency for enzymes to move up or down a substrate concentration gradient.  These have been seen for both fast exothermic enzymes like urease and slow endothermic enzymes like aldolase. These results  open a new perspective of enzymes as active matter \cite{jee_catalytic_2018}, and have potential practical and biological implications. For example, enzyme chemotaxis can be used to separate catalytically active and inactive enzymes \cite{dey_chemotactic_2014} and might contribute to the assembly of intracellular metabolons \cite{wu_krebs_2015,zhao_substrate-driven_2017} and to intracellular signaling \cite{weistuch_spatiotemporal_2018}. A number of mechanisms have been proposed for these phenomena, but their physical basis is still a subject of active research. 

Here, we analyze relationships among the various experimental and theoretical studies and discuss general theoretical considerations and frameworks for this field. (Related reviews also summarize recent work on enzyme EED and chemotaxis \cite{zhang_enhanced_2019,agudo-canalejo_enhanced_2018,astumian_enhanced_2014,dey_chemically_2017}.)  For EED and then chemotaxis, we review experimental findings and then analyze molecular mechanisms that have been proposed to account for these findings.  Central questions include how to explain observations of EED and chemotaxis in general; whether -- or when -- EED is an active process; and how to reconcile apparently inconsistent experimental observations. We conclude with a summary of key points and potential directions for future work to further elucidate these intriguing phenomena.

\section{EXPERIMENTAL STUDIES OF ENHANCED ENZYME DIFFUSION}
This section summarizes experimental reports of increases in enzyme diffusion coefficients, focusing on enzymes for which an active mechanism, such as execution of “swimming” motions by the enzyme, has been proposed.  We also discuss cases in which EED was expected but not observed, and consider potential experimental artifacts. 

\subsection{Positive observations}\label{exp. postive}
Apparent EED has been reported for varied enzymes, including ATPase \cite{borsch_conformational_1998}, T7 RNA polymerase \cite{yu_molecular_2009}, T4 DNA polymerase \cite{sengupta_dna_2014}, hexokinase \cite{zhao_substrate-driven_2017}, aldolase \cite{zhao_substrate-driven_2017}, alkaline phosphatase \cite{riedel_heat_2014}, acetylcholinesterase \cite{jee_enzyme_2018}, jack bean urease \cite{muddana_substrate_2010,sengupta_enzyme_2013,riedel_heat_2014,jee_catalytic_2018,jee_enzyme_2018,xu_direct_2019}, and catalase \cite{sengupta_enzyme_2013,riedel_heat_2014,jiang_observation_2017}. The relative diffusion enhancements, measured in homogeneous solution at the highest tested substrate concentrations, range from 15\% to 80\% (\textbf {Table \ref{tab1}}). Although these increases are similar, the turnover rates of these enzymes span orders of magnitudes (\textbf {Table \ref{tab1}}). This discrepancy argues against an active mechanisms as a general explanation of EED. Most of these studies used fluorescence correlation spectroscopy (FCS) which measures enzyme diffusion rates in homogeneous solution. However, one \cite{yu_molecular_2009} used fluorescence recovery after photobleaching (FRAP), and another used a relatively novel electrochemical method to support their FCS observation of EED \cite{jiang_observation_2017}. 

\begin{table}
\tabcolsep7.5pt
\caption{Turnover rates, hydrodynamic radii, minimum thrust speeds, and required reaction free energies, of enzymes reported to show EED. For details, see \cite{feng_thermodynamic_2019}. Citations are parenthesized.}
\label{tab1}
\begin{threeparttable}
\begin{tabular}{lcccc}
Enzyme & Turnover rate/\si{s^{-1}} & Radius/nm  & $v_{\text{min}}$/\si{m.s^{-1}} & \(-\Delta G^\circ_{\text{req}}\)/\si{kJ.mol^{-1}} \\
\hline
T4 DNA polymerase & 0.5\cite{burrows_purification_1992}&4.6\cite{challberg_purification_1979}  & \(1\times10^{-2}\) & \(1\times10^{7}\)\\
Aldolase & 5\cite{illien_exothermicity_2017} &4.9\cite{de_la_torre_calculation_2000} &  \(9\times10^{-3}\)&\(8\times10^{5}\)\\
T7 RNA polymerase & 4\cite{martin_kinetic_1987} &6.0\tnote{a} & \(6\times10^{-3}\) & \(6\times10^{5}\)\\
Hexokinase & 300\cite{zhao_substrate-driven_2017} &6.3\cite{lilie_yeast_2011}&  \(5\times10^{-3}\)& 6000 \\ 
ATP synthase & 1000\cite{iino_mechanism_2009}&6.6\cite{borsch_conformational_1998}& \(5\times10^{-3}\) &2000\\
Alkaline phosphatase & 3000\cite{riedel_heat_2014}&7.7\cite{ey_calf_1977} & \(3\times10^{-3}\)&  400\\ 
Catalase & 10000\cite{riedel_heat_2014} &5.3\cite{de_la_torre_calculation_2000} & \(7\times10^{-3}\) & 300\\ 
Urease & 10000\cite{riedel_heat_2014}&7.0\cite{follmer_jack_2004}& \(4\times10^{-3}\) &   100\\ 
Acetylcholinesterase & 20000\cite{froede_direct_1984}&8.8\cite{rieger_torpedo_1976}& \(3\times10^{-3}\) &  40\\ 

\hline
\end{tabular}
\begin{tablenotes}
\item[a] Calculated with HYDROPRO \cite{ortega_prediction_2011} from PDB structure 4RNP \cite{sousa_crystal_1993}.
\end{tablenotes}
\end{threeparttable}
\end{table}

Two FCS studies of Jee and coworkers \cite{jee_enzyme_2018,jee_catalytic_2018} provide FCS data at increased spatial and temporal resolution. By combining superresolution microscopy and FCS, these authors were able to reduce the horizontal diameter of the observation region – the beam waist -- to 50 nm, versus more typical values of about 800 nm \cite{muddana_substrate_2010}.  As the beam waist fell below $\sim$100 nm, an initially unimodal distribution of transit times was resolved into two peaks. The peak corresponding to longer transit times was attributed to conventional diffusion through the waist, and the peak corresponding to shorter times was attributed to fast “ballistic” motions induced by catalysis. 

Another study measured the diffusion of fluorescently labeled urease by single-molecule tracking with total internal reflectance fluorescence (TIRF) microscopy \cite{xu_direct_2019}. This method detects molecules only while they are about 300nm from the planar glass coverslip, so the enzyme molecules were confined in this layer by addition of methylcellulose to the solution.  This study yielded a remarkable 300\% increase in the diffusion coefficient upon addition of 1mM urea. Interestingly, the enhanced diffusion coefficient observed here, $\sim$ \si{3\times 10^{-13} m^2/s} is about 200-fold less than the non-enhanced value measured by FCS for urease in homogeneous solution \cite{sengupta_enzyme_2013}.  This drop in the diffusion coefficient might result from increased viscosity due to the methylcellulose. It is perhaps relevant that increasing the viscosity reduces the power required to achieve a given relative increment in the translational diffusion coefficient via self-propulsion (Section \ref{thermodynamics self-propulsion}, Eq \ref{Preq})\cite{feng_thermodynamic_2019}.  On the other hand, such a large change in the baseline diffusion coefficient raises the possibility that the physics of diffusion in this confined setting is significantly different from that in bulk solution.

\subsection{Evidence for and against an active mechanism of enhanced diffusion}\label{exp. active vs passive}
The concept that EED results from an active process has gained support from FCS and dynamic light scattering (DLS) studies indicating that catalytically active urease, catalase, and aldolase generate increased motion of passive tracer particles in solution \cite{zhao_enhanced_2017,jee_enzyme_2018}, with a magnitude and reaction-rate dependence similar to that of the enzymes’ own enhanced diffusion. Similarly, immobilized DNA polymerase, exhibits EED and generates fluid flow proportional to the catalytic rate \cite{sengupta_dna_2014}. The observation of a 20\% increase in the diffusion coefficient of passive tracers is particularly striking for aldolase, given that its turnover rate is only about \si{5/s} and that it was present at a concentration of 10 nM.   

If EED results from catalysis, then enzyme inhibitors should prevent EED. Accordingly, EED has not been observed in catalase in the presence of both substrate and and the inhibitor cyanide \cite{sengupta_enzyme_2013} or azide \cite{riedel_heat_2014}. However, at least three enzymes described as having catalysis-induced EED also have been reported to show EED in the absence of catalysis. First, Yu and coworkers \cite{yu_molecular_2009} reported that RNA polymerase has 25\% EED when catalytically active and 15\% EED when substrate is provided but by the required cofactor Mg2+ is withheld. Second, although heat release had been proposed as a mechanistic requirement for EED \cite{riedel_heat_2014}, Illien and coworkers \cite{illien_exothermicity_2017} later showed ~30\% EED for the enzyme aldolase, although this enzyme catalyzes an endothermic reaction \cite{zhao_substrate-driven_2017}. This led to the idea that catalysis itself might not be required for EED by aldolase, and the same authors used FCS to show that aldolase’s competitive inhibitor pyrophosphate could generate nearly the same level of EED as its substrate, fructose-1,6-bisphosphate (FBP). Third, urease did not show EED in the presence of the inhibitor pyrocatechol alone at 1mM concentration, but did show attenuated EED in the presence of both pyrocatechol and substrate \cite{muddana_substrate_2010}. Another study of urease found that a 1mM concentration of the substrate urea sufficed to cause EED, while the urease inhibitor boric acid began to cause EED only at higher concentrations (about 100 mM) \cite{jee_catalytic_2018}. The authors therefore proposed that substrate and inhibitor cause EED by two different mechanisms, to account for inconsistencies among experimental results as to when or whether EED requires catalysis.  

\subsection{Negative observations and possible experimental artifacts}\label{exp. negative}
Significant inconsistencies have emerged in studies of aldolase across multiple techniques. Although an FCS study \cite{illien_exothermicity_2017} indicated about 30\% EED in the presence of either substrate or a competitive inhibitor, Zhang and coworkers \cite{zhang_aldolase_2018} studying aldolase using DLS found no EED in the presence of either substrate or inhibitor. Gunther and coworkers \cite{gunther_absolute_2019} also observed no EED when studying aldolase by a third technique, DOSY NMR. It is not yet clear how to reconcile all of these results, but another study from Gunther and coworkers \cite{gunther_diffusion_2018} highlights potential artifacts and complexities of the widely used FCS technique. 

One potential source of error in FCS is that there is always some free fluorophore, so, if more protein binds to the glass over time, the relative contribution of the fast-diffusing fluorophore to the measured diffusion coefficient increases, leading to an artifactual increase in the apparent diffusion coefficient \cite{gunther_diffusion_2018}. However, results of Illien and coworkers \cite{illien_exothermicity_2017} argue against this artifact, because they found that the elevated diffusion coefficient of aldolase in the presence of substrate returned to baseline once the substrate was consumed. 

Quenching of the fluorophore by substrate or product could also lead to errors, as early suggested by Bai and Wolynes \cite{bai_hydrodynamics_2015}. Indeed, Gunther and coworkers \cite{gunther_diffusion_2018} showed that this artifact can account for apparent EED of the enzyme alkaline phosphatase \cite{riedel_heat_2014} when the quenching substrate nitrophenyl phosphate is used, because the apparent EED disappears when a nonquenching substrate is used. Given that EED in catalase has been studied by FCS, it is worth noting that its substrate, hydrogen peroxide, also can act as a quencher \cite{gunther_diffusion_2018}.  On the other hand, an FCS study of EED in urease argues against the quenching artifact by confirming that urea does not reduce the fluorescence lifetime of the fluorescent label \cite{jee_catalytic_2018}. 

Finally, FCS measurements are typically carried out at enzyme concentrations roughly 1000 times lower than DLS and DOSY measurements \cite{gunther_diffusion_2018,zhang_aldolase_2018}. This makes it more probable that some of the multimers have dissociated in the FCS studies. Given that binding of substrate and/or inhibitor molecules sometimes promotes dissociation  \cite{borsch_conformational_1998,woodfin_substrate-induced_1967}, binding could increase the measured diffusion coefficient merely by causing enzymes to dissociate into faster-diffusing subunits. In any case, because most reports of EED, for all enzymes studied, rely exclusively on FCS, it would seem important to track down the cause of the discrepancy for aldolase and/or to apply alternative experimental methods to other enzymes.

\section{POTENTIAL MECHANISMS OF ENHANCED ENZYME DIFFUSION}\label{mechanisms eed}
Here we analyze proposed mechanistic explanations of EED. These are divided in two categories, propulsive and non-propulsive. Before discussing specific mechanisms, we consider the thermodynamics of active self-propulsion and the possibility of a role for hydrodynamic interactions among enzymes in solution. It is worth emphasizing at the outset that, to be plausible, a mechanism must meet both qualitative and quantitative criteria. That is, the proposed mechanism must not only be physically workable but also capable of generating EED at the levels observed experimentally when realistic values of parameters, such as $k_{cat}$ and hydrodynamic radius, are considered. For example, in principle, an enzyme’s diffusion coefficient could rise due to heating of the solution by the enzyme-catalyzed reaction.  However, under normal conditions, the temperature does not rise nearly enough to account for observed levels of EED (Section \ref{temperature increase}).

\subsection{A thermodynamic constraint on enzyme self-propulsion}\label{thermodynamics self-propulsion}

Varied physical mechanisms have been proposed by which chemical energy released via enzyme catalysis could lead to propulsion, thus increasing an enzyme's translational diffusion coefficient. Any propulsion mechanism necessarily leads to dissipation of chemical energy, such as by viscous drag opposing the propelled motion.  Therefore, the entire class of self-propulsion mechanisms is plausible only when the catalyzed reaction provides enough power to match the unavoidable dissipation. We have recently analyzed this generally applicable limit, quantitatively connecting theory with experimental data \cite{feng_thermodynamic_2019}, as now summarized.

A fundamental aspect of self-propulsion mechanisms is that the enzyme is an asymmetric particle, which is considered to be propelled in a given direction within its own frame of reference.  As a consequence, the enzyme's rotational diffusion causes its lab-frame propulsion direction to change stochastically, according to the rotational diffusion constant $D_r$. The overall translational diffusion of such a self-propelled particle results from normal Brownian motion combined with this randomly oriented self-propulsion. Using the Stokes-Einstein law, one can write an apparent diffusion coefficient as the sum of the non-enhanced translational diffusion coefficient $D_t$ and a contribution from propulsion with speed $v$:

\begin{equation}\label{Dapp}
D_{app} =  D_t + \frac{v^2}{6D_r}\frac{t_p}{t_c}
\end{equation}

where $\frac{t_p}{t_c}$ is the mean fraction of the catalytic cycle during which the propulsive force is present. The minimal power required to achieve a certain ratio of diffusion enhancement, $R = \frac{D_{app}}{D_t} -1$,  is given by the Stokes drag dissipation rate \cite{lighthill_squirming_1952,stone_propulsion_1996,sabass_efficiency_2010,sabass_dynamics_2012,wang_chemical_2005}:

\begin{equation}
\label{Preq}P_{\text{req}} = 6\pi\eta a v^2 \frac{t_p}{t_c} = \frac{3}{4}\frac{R(kT)^2}{\pi\eta a^3}
\end{equation}

where $a$ is the hydrodynamic radius and $\eta$ is viscosity. Using the hydrodynamic radii and turnover numbers of enzymes reported to undergo EED allows one to estimate the free energy per reaction required to generate experimental levels of diffusion enhancement \cite{feng_thermodynamic_2019}. The required free energies (\textbf {Table \ref{tab1}}) are orders of magnitude larger than available for the slower enzymes and considerably larger than available even for the fastest enzymes, despite the use of conservative assumptions in the power analysis \cite{feng_thermodynamic_2019}. 

The strong inverse dependence of $P_{req}$ on hydrodynamic radius $a$ helps explain how self-propulsion can still lead to significant enhanced diffusion for larger particles with asymmetrically disposed catalytic sites, such as so-called Janus nanoparticles \cite{lee_self-propelling_2014}. Intuitively, the orders-of-magnitude slower rotational diffusion of these larger particles makes the self-propulsion trajectory much less tortuous and more effective at generating net displacement from an initial location. 

As part of this analysis, we considered the chemical free energy available to power propulsion.  In principle, the chemical power available to drive propulsion is the product of the reaction rate and the free energy change of the reaction under the experimental conditions studied. This free energy change depends on the concentrations of reactants and products.  However, global concentrations do not couple to the local processes of substrate binding, chemical reaction, and product release, so the global free energy of reaction may not be what is available to drive propulsion; it may instead be a local free energy \cite{anderson_colloid_1989} that is relevant.  We used a thermodynamic cycle to estimate local free energy changes of enzyme-catalyzed reactions and found that, fortuitously, these are generally close to the standard reaction free energies \cite{feng_thermodynamic_2019}. Experimental measurements of enzyme diffusion when the reaction is at equilibrium or running in reverse might test these ideas and shed light more generally on mechanisms of EED.

\subsection{The insignificance of hydrodynamic interactions}\label{hydrodynamic interaction}
The thermodynamic analysis in Section \ref{thermodynamics self-propulsion} assumes that the enzyme molecules in solution move independently. However, EED might be amplified if the motions of each enzyme molecule could be further driven by the motions of others. Because the enzyme solutions used in FCS measurements are dilute, typically 10nM, the average distance between two enzyme molecules is about 550nm. At this range, any intermolecular forces are extremely weak. For example, the Coulombic interaction between two enzymes of charge +10, assuming water’s dielectric constant, is less than 0.1 \si{kcal.mol^{-1}}, even neglecting ionic screening. 
However, hydrodynamic interactions among enzyme molecules might provide a mechanism for longer-ranged enzyme-enzyme correlations. Indeed, several studies have indicated that enzyme activity, including that of aldolase, can generate fluid flows \cite{sengupta_dna_2014} or enhanced diffusion of passive tracer molecules\cite{zhao_enhanced_2017}, possibly via hydrodynamic interactions. In addition, Sengupta and coworkers observed no increase in the diffusion coefficient of fluorescently labelled, inactivated catalysis in the presence of unlabeled, active catalase \cite{sengupta_enzyme_2013}.

A theoretical study has shows that hydrodynamic interactions are proportional to the concentration of active enzyme, and it was argued that hydrodynamic interactions among enzymes could lead to significant increments in diffusion coefficients \cite{mikhailov_hydrodynamic_2015}. However, their numerical calculation assumed an enzyme concentration of 1000 nM, far higher than those used in FCS measurements. If the typical concentration 10nM is used, the predicted enhancement in diffusion coefficient by hydrodynamic interactions comes to only $6\times10^{-13}\si{m^2/s}$, much less than reported values of EED. From another perspective, the flow velocity field generated by an enzyme decays no slower than \(r^{-2}\) \cite{elgeti_physics_2015,mikhailov_hydrodynamic_2015}. Thus, in a 10nM solution, the flow velocity generated near the surface of one enzyme (radius $\sim$10nm) will have decayed at least about 3000-fold at the position of another enzyme (distance $\sim$ 550nm). Thus, it seems unlikely that hydrodynamic interactions contribute significantly to EED. Future measurements examining the magnitude of EED as a function of enzyme concentration might offer further insight. 

\subsection{Self-propulsion mechanisms}\label{self-propulsion mechanisms}
 
\subsubsection{Mechanical swimming}\label{mechanical swimming}
By mechanical swimming, we mean the generation of propulsion by a repeated cycle of conformational changes. Microorganisms engage in mechanical swimming \cite{lauga_hydrodynamics_2009,elgeti_physics_2015}, but swimming by non-motor enzymes is not well established. The scallop theorem implies that a simple cycle of forward and reverse conformational changes cannot generate net propulsion, as any motion induced by the forward step will be undone by the reverse step \cite{lauga_life_2011}. However, the chiral character of enzymes means that motions driven by an out-of-equilibrium chemical reaction will be directional and hysteretic \cite{slochower_motor-like_2018}, and thus capable of generating net propulsion \cite{slochower_motor-like_2018}. Furthermore, even if the conformational changes were perfectly time-reversable, they could generate an increase in the translational diffusion coefficient, because rotational diffusion of the enzyme during the enzymatic cycle allows the motion generated by the forward step to be along a different lab-frame axis than the motion generated by the reverse step \cite{lauga_enhanced_2011}. Nonetheless, mechanical swimming is an unlikely explanation for EED, based on the thermodynamic argument provided in Section \ref{thermodynamics self-propulsion}, and on another study indicating that any plausible enzyme motions are too weak to generate observed levels of EED \cite{bai_hydrodynamics_2015}. 

\subsubsection{Pressure waves}\label{pressure waves}
Riedel and coworkers suggested that the rapid release of heat at the catalytic step of an exothermic, enzyme-catalyzed reaction could generate a pressure wave that produces an asymmetric force on the enzyme, leading to self-propulsion.  Bai and Wolynes \cite{bai_hydrodynamics_2015} argued against this mechanism by showing that an extremely large conformational motion, along the lines of complete unfolding and refolding, would be needed.  Our interpretation of the original suggestion is that the pressure wave comes not from a fast conformational change, but from sudden heating at the catalytic site due to the chemical reaction. This view might avoid the concern raised by Bai and Wolynes, but it would still be unclear how passage of a single pressure pulse through the enzyme at the speed of sound could generate a large net displacement of the enzyme. (Note that the passage of sound waves through water or air does not lead to net displacement of the molecules forming the medium.) Golestanian \cite{golestanian_enhanced_2015} has also provided a theoretical argument against this proposed mechanism.

\subsubsection{Bubble propulsion}\label{bubble}
The enzyme catalase has molecular oxygen as a product, and a sufficiently high density of catalase molecules on a surface can generate oxygen bubbles, leading to a propulsive force \cite{dey_chemically_2017,pantarotto_autonomous_2008,sanchez_dynamics_2010}. However, generation of bubbles was, arguably, ruled out as a mechanism for self-propulsion of catalase by direct observation and by demonstration that active catalase does not increase the diffusion coefficient of nearby passive tracer molecules \cite{sengupta_enzyme_2013}. Also, most of the enzymes for which EED has been reported do not create a potentially gaseous product.

\subsubsection{Phoretic self-propulsion}\label{phoretic}
Phoretic mechanisms play an important role in self-propelled synthetic particles, such as Janus nanomotors \cite{santiago_nanoscale_2018}, and the underlying theory underlying is well-developed \cite{moran_phoretic_2017}. Self-phoretic mechanisms include self-diffusiophoresis, self-electrophoresis, and self-thermophoresis, which result, respectively, from interactions of a particle with self-induced gradients of concentration, electrical potential, or temperature. These mechanisms are unlikely to explain EED, because the thermodynamic limit on self-propulsion discussed above applies to phoretic self-propulsion. In addition to frictional dissipation, phoretic self-propulsion would require extra power to maintain the self-induced gradient \cite{sabass_dynamics_2012}, further increasing the gap between the required power and the power available from the chemical reaction. Nonetheless, it is informative to consider specific phoretic mechanisms that have been put forward.

Self-electrophoresis has been suggested as a mechanism of EED \cite{muddana_substrate_2010}, but seems unlikely. First, EED has been reported for enzymes whose substrates and products are electrically neutral, so they cannot set up an electric field. Second, self-electrophoresis would be influenced by ionic strength, but ionic strength was reported to have no influence on the apparent EED of RNA polymerase \cite{yu_molecular_2009}. Third, although self-electrophoresis has been reported for Janus nanomotors, these have spatially separated ionic flows at their their "cathodes" and "anodes", and are thus well suited to create ionic gradients and resulting electrical fields. In contrast, enzymes usually bind substrate and expel product at the same site, and therefore are not as good at generating gradients. 

Arguing in favor of a self-diffusiophoresis mechanism, Colberg and Kapral \cite{colberg_angstrom-scale_2014} presented simulations of enzyme-sized particles undergoing diffusiophoresis at high propulsion speeds of roughly 4\si{m/s}.  However, the study assumes a diffusion-controlled reaction with a rate constant of about \si{4\times 10^{10}/M.s}, which is much larger than enzyme turnover rates listed in \textbf {Table \ref{tab1}}. It also assumes a concentration of substrate much higher than that in enzyme systems -- the substrate was essentially a solvent —-so the diffusiophoretic forces could be unrealistically large. Finally, the magnitude of the attractive forces between the enzyme and substrate and product are arbitrary, rather than being chosen to reflect typical enzyme-substrate interactions. Thus, their model does not closely resemble an enzyme-substrate system and is of limited applicability.

Finally, Golestanian estimated the magnitude of the potential self-thermophoretic self propulsion in catalase and found it negligible \cite{golestanian_enhanced_2015}; and self-thermophoresis should be even weaker for other enzymes that release less heat.

\subsection{Non-propulsive mechanisms}\label{Non-propulsive mechanisms}
We now consider mechanisms of EED that do not involve self-propulsion and therefore avoid the concerns raised above regarding the inadequacy of the chemical power available. 

\subsubsection{Local pH change}
Muddana and coworkers \cite{muddana_substrate_2010} considered whether EED could, for some enzymes at least, result from local changes in pH due to enzymatic activity. For example, ammonia, one of the products of the urease reaction, is basic and therefore can change the pH.  However, measurement of the pH in the immediate vicinity of urease with a pH-sensitive fluorophore covalently bound to the enzyme revealed pH increases of up to only $\sim$ 0.8, which were judged insufficient to explain EED. Note, too, that many enzymes for which EED has been reported cannot change the pH.

\subsubsection{Temperature increase}\label{temperature increase}
Another way for catalysis to increase the diffusion constant of an enzyme would be for the released heat (if any) to increase the temperature of the solution. However, even for an exothermic enzyme-catalyzed reaction, the increase in the bulk temperature of the solution is far too small to account for observed EED \cite{riedel_heat_2014,golestanian_enhanced_2015,tsekouras_comment_2016}. Golestanian \cite{golestanian_enhanced_2015} has proposed a more refined mechanism, "collective heating", which accounts for the nonequilibrium heat flow in the measuring container and for the increase in enzyme turnover with increasing temperature, and has argued that these factors can lead to a large enough temperature increase and viscosity decrease to account for EED.  However, this study uses the thermal conductivity of air in its numerical analysis, and when the 30-fold higher value of water is used instead, as would seem appropriate, the predicted temperature increase no longer appears sufficient.

\subsubsection{Changes in conformation, conformational fluctuations, and quaternary structure}
If binding of substrate generates a new conformational distribution of the enzyme with a smaller mean hydrodynamic radius, this would lead to EED. In addition, the degree of EED would correlate with the enzyme’s catalytic rate, as often observed, because increasing substrate concentration will both increase the catalytic rate and the fraction of bound enzyme. This explanation is appealing because it avoids the power requirements of propulsive mechanisms. It also could account for observations of EED induced by binding of an inhibitor, or by addition of substrate to an enzyme that lacks an essential cofactor. However, experimental data compiled from the literature \cite{zhang_enhanced_2019} and Brownian dynamics simulations \cite{kondrat_brownian_2019} suggest that binding of a substrate or an inhibitor does not cause a large enough reduction in an enzyme's mean radius of gyration to account for observed levels of EED.

Recently, Illien and coworkers showed theoretically that the diffusion coefficient can be increased by a decrease in the thermal fluctuations of the particle radius \cite{illien_exothermicity_2017,illien_diffusion_2017}. They note that an enzyme may be stiffened by binding of another molecule, so addition of substrate or inhibitor could lead to concentration-dependent EED, as observed experimentally. However, the Brownian dynamics study mentioned in the prior paragraph \cite{kondrat_brownian_2019} suggests that this mechanism would not lead to observed levels of EED. 

Finally, Gunther and coworkers \cite{gunther_diffusion_2018} have pointed out that many of the enzymes for which EED has been reported are multimeric, and that binding of substrate can lead to dissociation of multimeric enzymes. Because dissociation into smaller components would lead to an increase in the diffusion coefficient, averaged over the various multimers in solution, binding-induced dissociation could provide another non-propulsive explanation for observations of EED. This mechanism may be particularly relevant for FCS measurements, which are typically run at low enzyme concentrations that shift the equilibrium toward dissociated states \cite{lauga_life_2011}. Dimeric yeast hexokinase, for which EED has been observed via FCS \cite{zhao_substrate-driven_2017}, has a dissociation constant of 0.1-1.0 $\mu M$, which is well above 10nM level concentrations typically used in FCS experiments. Furthermore, three enzymes for which EED has been observed, hexokinase, urease, and acetylcholinesterase, were reported to dissociate as substrate concentrations rise above their respective $k_M$ values \cite{jee_enhanced_2019}. It is thus worth noting that, if different-sized multimers interconvert on an appropriate timescale (tens of microseconds), this would cause an additional decay mode in the FCS autocorrelation function, and thus could offer an alternative explanation for the appearance of a bimodal distribution of transit times in high-resolution FCS studies \cite{jee_catalytic_2018,jee_enzyme_2018}. However, dissociation does not seem like a viable explanation for at least some observations of EED, because the tetramer-to-dimer dissociation constant of aldolase is about 1 pM \cite{zhang_aldolase_2018}, so this enzyme should be quite stable as a multimer in most or all of the relevant experiments; and the single-molecule tracking study of urease by Xu and coworkers \cite{xu_direct_2019} reported no influence of the concentrations of urea or enzyme on dissociation or EED. In addition, there is evidence of EED by urease at substrate concentrations where little dissociation is evident \cite{jee_enhanced_2019}.  

\section{EXPERIMENTAL OBSERVATIONS OF ENZYME CHEMOTAXIS}\label{exp. chemotaxis}
Motile bacteria and eukaryotic cells have evolved mechanisms for swimming up or down gradients of dissolved compounds.  This directed movement of cells is called chemotaxis, and movement up or down a gradient is termed attractive or repulsive chemotaxis, respectively.  Bacteria use a molecular memory system to determine whether their recent motions have taken them up or down the gradient and adapt accordingly, while eukaryotic cells use their size to sense the direction of a gradient across the cell in real time \cite{schnitzer_strategies_1990}. In recent years, it has been reported that enzymes also can move preferentially up or down a substrate gradient \cite{sengupta_enzyme_2013,jee_enzyme_2018}, with  apparent drift speeds up to about \si{1\mu m/s} for attractive enzyme chemotaxis \cite{sengupta_enzyme_2013}, and \si{10\mu m/s} for repulsive chemotaxis \cite{jee_enzyme_2018}.

Enzymes cannot meet the strictest definition of chemotaxis \cite{wheat_collective_2011,wilkinson_assays_1998} because they lack the memory and/or size required to mimic either the bacterial or eukaryotic mechanisms. Instead the apparent directional migration of enzymes may arise from factors such as space-dependent enzyme diffusivity or diffusiophoresis, as discussed in Section \ref{mechanism chemotaxis}. In this paper, we follow the literature in applying the term chemotaxis to all observations of enzymes moving preferentially up or down a concentration gradient. 

\subsection{Attractive chemotaxis}\label{attractive chemotaxis}
\begin{figure}
\includegraphics[width=6in]{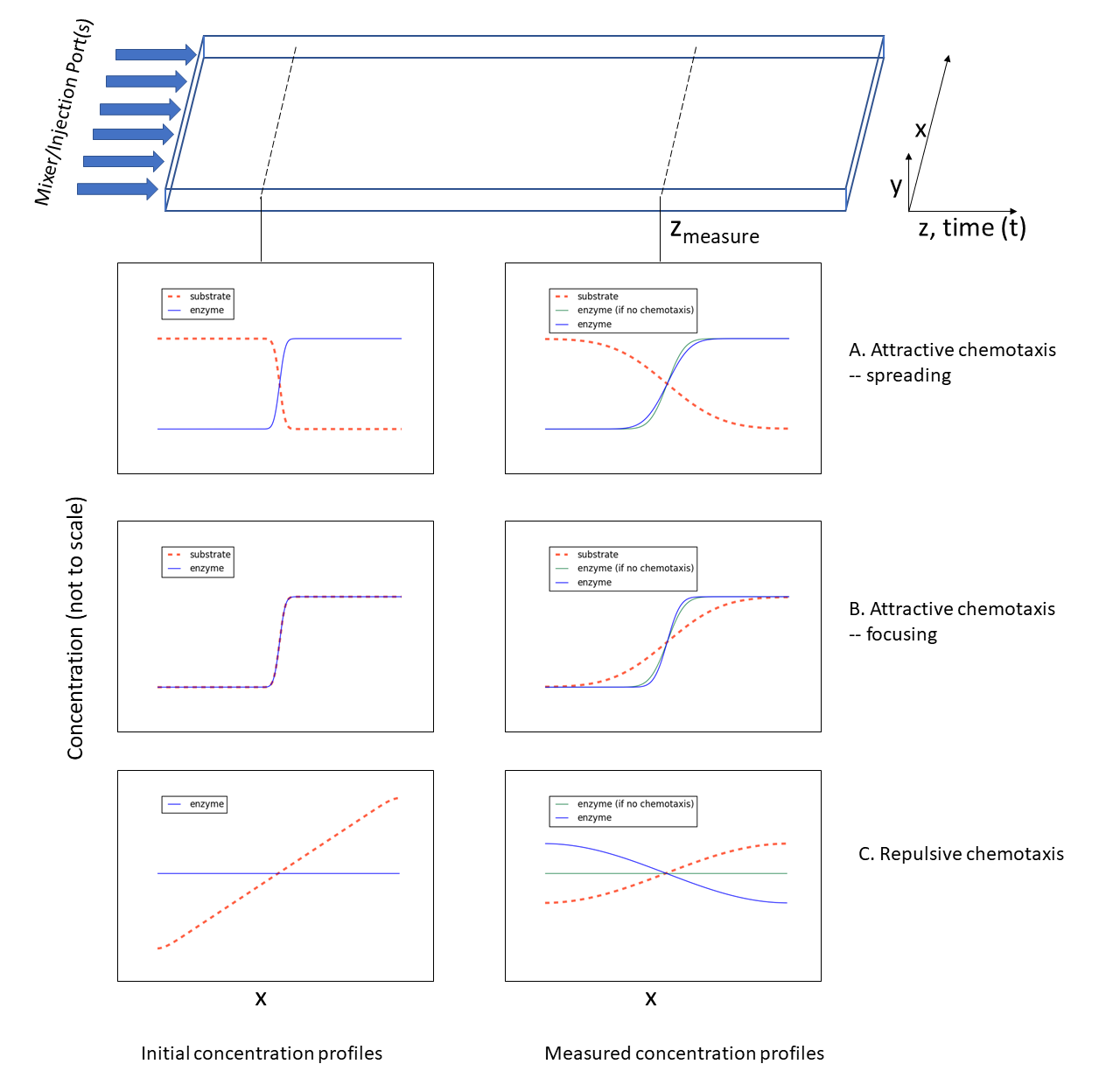}
\caption{Generic diagram of a microfluidic chemotaxis experiment, with laminar flow from left to right; i.e., in the +z direction.  Concentration profiles of enzyme (solid blue or green) and substrate (dashed red) across the channel (x-axis) are initialized by injection ports and mixers (initial concentration profiles in plot) and measured (measured concentration profiles in plot) after the concentration profiles have relaxed for some time t, corresponding to distance $z_\text{measure}$ from the injection ports. A. Concentration profiles corresponding to a transient “lurch” of enzyme toward substrate, suggestive of attractive chemotaxis \cite{sengupta_enzyme_2013}. B. Focusing of enzyme in the high-substrate part of the channel, suggestive of attractive chemotaxis \cite{zhao_substrate-driven_2017}. C. repulsive chemotaxis in the context of an initially linear substrate profile \cite{jee_catalytic_2018}.}
\label{fig1}
\end{figure}

An early report of attractive enzyme chemotaxis used fluorescence to detect preferential displacement of RNA polymerase up a gradient of its NTP substrate in a millimeter-scale device \cite{yu_molecular_2009}. Subsequent experiments have measured fluorescence intensity profiles of labelled enzymes across microfluidic flow channels fed by incoming channels containing solutions with differing compositions (\textbf{Figure \ref{fig1}}). For example, a left feed might contain either plain buffer or substrate in buffer and a right feed might contain enzyme in buffer; here attractive chemotaxis could manifest by a tendency of the enzyme to move to the left more in the presence of substrate than in its absence (\textbf{Figure \ref{fig1}}A). Because flow in microfluidic devices is laminar, diffusion, rather than convection, dominates the relaxation of the initial non-equilibrium concentration. Such devices showed preferential diffusion of catalase and urease toward their respective substrates \cite{sengupta_enzyme_2013,dey_chemotactic_2014}, and of DNA polymerase toward either its substrate or its Mg2+ cofactor in the presence of substrate \cite{sengupta_dna_2014}.  A technically similar study of mitochondrial malate dehydrogenase and citrate synthase showed chemotaxis toward substrate, both in the presence and absence of required enzyme cofactors \cite{wu_krebs_2015}. Thus, catalysis is not always necessary for enzyme chemotaxis to be observed.

A potential weakness of studies like those in the prior paragraph is that enhanced movement of enzyme into the region of the microfluidic channel with substrate might result only from faster diffusion in the presence of substrate, rather than from any directional preference \cite{jee_catalytic_2018}.  Studies by Sen and coworkers address this. When hexokinase and its substrates D-glucose and ATP were injected into the center of the channel, with plain buffer on both sides, the spread of the active enzyme across the channel was slower than when the same experiment was done with inactive enzyme \cite{zhao_substrate-driven_2017,mohajerani_theory_2018}. In effect, the enzyme was retained in the central channel. This “focusing” result implies a real tendency of the enzyme to remain close to the substrate. In addition, catalysis appears to be important in this effect, because it was attenuated when mannose, a substrate with slower turnover was used instead of D-glucose \cite{zhao_substrate-driven_2017,mohajerani_theory_2018}, and it was absent with D-glucose but no ATP, and with ATP but L-glucose, which is not a substrate  \cite{mohajerani_theory_2018}. 

\subsection{Repulsive chemotaxis}\label{repulsive chemotaxis}
Paradoxically, Jee and coworkers \cite{jee_enzyme_2018}, also using a microfluidic device, reported repulsive chemotaxis for urease and acetylcholinesterase, as an initially uniform enzyme concentration increased on the side of the channel with lower substrate concentration. Agudo-Canalejo et al. have suggested that this difference, relative to prior experiments, might result from the use of a different range of substrate concentrations \cite{agudo-canalejo_phoresis_2018}, but the observation was later confirmed over a wider concentration range \cite{jee_catalytic_2018,mohajerani_theory_2018}. It may also be relevant that Jee and coworkers used a different experimental design, in which the enzyme was initially at uniform concentration across the channel \textbf{Figure \ref{fig1}}. In addition, it appears that Jee and coworkers’ measurements \cite{jee_enzyme_2018} allowed  more interaction time (distance from start of channel $\times$channel area / fluid flow rate) than the studies that observed attractive chemotaxis \cite{sengupta_enzyme_2013,mohajerani_theory_2018}. Thus, the experiments of Jee and coworkers might reveal something closer to the shape of the ultimate steady state distribution of the enzyme.
 
\subsection{The microfluidic method}\label{microfluidic}

Two important features of the microfluidic experiments are worth highlighting. First, they do not give the steady state distribution of an enzyme in the context of a time-invariant substrate gradient, because the concentration profiles of not only the enzyme but also the substrate change as the fluid progresses along the channel.
Instead, they report a transient response to a time-varying substrate gradient. Moreover, the diffusion coefficient of the substrate is usually much higher than that of the enzyme, so the concentration profile of the substrate normally relaxes more quickly than that of the enzyme. This is important because some physical mechanisms can explain transient movement of the enzyme along a substrate gradient but lead to a uniform distribution at steady state under a time-invariant substrate gradient. The microfluidic experiments may reveal the transient response but not the steady-state response.  It is perhaps worth mentioning that a microfluidic setup can, in another sense, be said to have reached its steady state after enough wall clock time has passed for its flows and concentration profiles to have stabilized. This is different from the question of whether one has taken a meaurement far enough down the channel -- i.e. long enough after mixing has begun -- for the concentration profiles to have stabilized as a function of the mixing time. Second, when interpreting the concentration distribution of enzyme across the channel, it should be borne in mind that that laminar flow has a parabolic profile, so detailed interpretation requires accounting for the details of fluid flow by simulations, as previously done \cite{mohajerani_theory_2018}. 

\section{POTENTIAL MECHANISMS OF ENZYME CHEMOTAXIS}\label{mechanism chemotaxis}
Potential mechanisms of enzyme chemotaxis may be grouped into two categories. One posits a force directed parallel to the substrate (or inhibitor) concentration gradient that drives either attractive or repulsive chemotaxis. The other is based on the idea that the enzyme's diffusion coefficient depends on the substrate concentration, leading to a position-dependent diffusion coefficient when there is a substrate concentration gradient.   Importantly, as emphasized by Agudo-Canalejo and coworkers, both categories of mechanisms can be at work in the same system \cite{agudo-canalejo_phoresis_2018}. Before considering specific mechanisms, however, we present a theoretical framework for defining and analyzing proposed mechanisms of enzyme chemotaxis.

\subsection{A theoretical framework for mechanisms of enzyme chemotaxis}\label{theoretical framework}
Enzymes translate through solvent in a Brownian manner and can thus be modeled by the Fokker-Planck (FP) equation \cite{zwanzig_nonequilibrium_2001}. This equation describes how an initial probability distribution (i.e. concentration) in space $p(x,t=0)$, evolves over time, and we have simplified by considering a one-dimensional system. For example, if $x$ represents the distance of a point from one edge of a microfluidic channel (\textbf{Figure \ref{fig1}}), an initial step-function concentration profile across the inlet end of the channel decays over time to a sigmoid-like distribution and approaches uniformity as the flow progresses further down the channel. The FP equation obeys conservation of probability, which requires that $\frac{\partial p}{\partial t}=-\frac{\partial }{\partial x}J$, where $J=J(x,t)$ is the probability flux. If the diffusion coefficient, $D$, is constant, then the flux can be written as $J=\mu Fp - D\frac{\partial p}{\partial x}$, where the first term represents directional drift induced by a force $F$, and $\mu$ is the enzyme's mobility (the reciprocal of friction coefficient). 

If an enzyme diffuses at a different rate in the presence of substrate, such as by EED, then the enzyme’s diffusion coefficient will depend on position, so $D=D(x)$. (The diffusion coefficient could also be a function of time, $D=D(x,t)$, because the substrate gradient in a microfluidic device decays with time (Section \ref{microfluidic}). However, the mathematical consequences of this complication are not considered in current literature on enzyme chemotaxis, and thus are not considered here.) Perhaps surprisingly, merely specifying $D(x)$ does not fully determine the correct form of FP equation. This is because the flux expression depends on the character of the microscopic process that causes $D(x)$ to vary with $x$ - an issue known as the Ito-Stratonovich dilemma, which arises in systems with multiplicative noise \cite{landauer_stability_1983,sokolov_ito_2010,farago_fluctuationdissipation_2014,tupper_paradox_2012}. A range of scenarios is captured by the following expression 
$J=\mu Fp-\alpha \frac{\partial D(x)}{\partial x}p-D(x)\frac{\partial p}{\partial x}$, where $0<\alpha <1$. This leads to the following relatively general form of the FP equation:

\begin{equation}
    \frac{\partial p}{\partial t}=-\frac{\partial }{\partial x} \left( \mu F p \right) +\alpha \frac{\partial}{\partial x}\left(p\frac{\partial D(x)}{\partial x}\right)+\frac{\partial}{\partial x}\left(D(x)\frac{\partial p}{\partial x}\right)
\end{equation}

Here, $\alpha \frac{\partial D(x)}{\partial x}p$ is an additional term that could contribute to chemotaxis.
In particular, if $\alpha=1$, then the steady state distribution of enzyme will be greater where the diffusion coefficient is lower \cite{jee_catalytic_2018,agudo-canalejo_phoresis_2018}. In contrast, if $\alpha=0$, the steady state distribution will be uniform in the absence of a force $F$ \cite{tupper_paradox_2012,jee_catalytic_2018}. The FP equation with $\alpha=1$ is often termed its Ito form, while the FP equation with $\alpha=0$ is termed the isothermal form \cite{sokolov_ito_2010,farago_fluctuationdissipation_2014}, and we will use these names below.

A fundamental conclusion of this analysis is that enzyme chemotaxis cannot be mechanistically explained by merely positing a position-dependent diffusion coefficient induced by a substrate gradient. This is because the drift of the enzyme depends on $\alpha$, which depends on the microscopic origin of the position-dependence of $D(x)$. Nonetheless, the FP equation is a valuable framework for understanding diffusive motion, and and the following section uses it to consider various possible mechanisms of enzyme chemotaxis. 

\subsection{Mechanisms based on force-induced drift}
If a substrate gradient leads to a net force, $F$, on enzyme molecules in solution, this will induce drift either up (attractive) or down (repulsive) the gradient, which could account for experimental observations of enzyme chemotaxis. This class of mechanism could explain the enzyme focusing result of Sen and coworkers \cite{zhao_substrate-driven_2017,mohajerani_theory_2018}, and the evolution of an initially uniform enzyme profile into a nonuniform one in the presence of a substrate gradient \cite{jee_catalytic_2018}. Force-induced drift mechanisms do not require a space-dependent diffusion coefficient, so multiplicative noise is not an issue. Two specific proposals for mechanisms in this class are now considered.

\subsubsection{Thermodynamic force}\label{thermodynamic force}
One mechanism derives an expression for a time-averaged force on macromolecules arising from the thermodynamics of macromolecule-cosolute binding in the presence of a concentration gradient of the cosolute. Schurr and coworkers included this concept in their analysis of chemotaxis of non-enzyme molecules \cite{schurr_theory_2013}, and good correlation with experiment was obtained for a case of enzyme chemotaxis \cite{zhao_substrate-driven_2017}. Mohajerani and coworkers modified this theory for enzymes, arguing that catalysis-associated EED could magnify the effect by increasing the baseline diffusion coefficient of the enzyme and thus the speed of chemotaxis, and they reported agreement with experimental data \cite{mohajerani_theory_2018}. The fundamental picture of this model is that a free enzyme molecule tends to move in the direction of higher substrate concentration when it binds, whereas an enzyme-substrate complex does not have any directional preference. 

A concern with this proposed mechanism is that it equates the position-averaged force on the enzyme, which is computed from the thermodynamic gradient, with the time-averaged force on the enzyme, which is the quantity relevant for chemotaxis. It is the time-averaged force that is relevant, because the instantaneous velocity of an overdamped enzyme molecule is proportional to the instantaneous force, and we are interested in the mean of this velocity over time. If the free energy fell linearly with position, the time-averaged and space-averaged forces would be equal. However, the free energy falls only at moments when the enzyme is transitioning from the unbound state to the substrate-bound state. These transitions are short-lived, because the binding forces are short-ranged, and these brief transitions are separated by long time intervals during which the enzyme feels no directing force. The lengths of these intervals are governed by the association and dissociation rate constants. Thus, an enzyme diffusing in solution feels the binding force for only a small fraction of the time, and the time-averaged force on the enzyme is expected to be far smaller than the position-averaged force, so it is not clear how well this proposed mechanism can account for enzyme chemotaxis. This reasoning appears analogous to that in a prior study showing that the Stokes efficiency of a molecular motor is less than one and may be very low indeed when the driving potential is not linear in the spatial variable along which the motor moves \cite{wang_stokes_2002}.  

Agudo-Canalejo and coworkers have offered additional points of concern regarding the thermodynamic model \cite{agudo-canalejo_enhanced_2018}, such as the fact that it cannot account for observations of repulsive chemotaxis (Section \ref{repulsive chemotaxis}). The possibility of further complexities with a mechanism based on the thermodynamics of enzyme-substrate binding is suggested by a prior observation that thermodynamics alone does not determine the phoretic speeds, or even the directions, of colloid particles (page 94 of \cite{anderson_colloid_1989}). More broadly, determining molecular motions requires knowing more than a thermodynamic tendency, i.e. an energy gradient; one must also know how chemical coordinates and mechanical coordinates are coupled to transduce this energy \cite{wang_ratchets_2002,astumian_physics_2016}.

\subsubsection{Diffusiophoresis}\label{diffusiophoresis}
Diffusiophoresis causes directional drift of colloid particles up or down the concentration gradient of a cosolute \cite{anderson_colloid_1989,khair_diffusiophoresis_2013}. It results from net attractive or repulsive forces between the particles and the cosolutes, leading to attractive  or repulsive chemotaxis. Diffusiophoresis is similar to self-diffusiophoresis (Section \ref{phoretic}) except that the gradient is externally imposed instead of being self-generated by catalysis.

The particle-cosolute interactions that cause diffusiophoresis are typically nonspecific, often involve long-ranged electrostatics, and may be averaged across the surface of a relatively large colloid particle \cite{schurr_theory_2013,moran_phoretic_2017}. We are not aware of studies deriving diffusiophoretic velocities for binding of an enzyme to its substrate, but Agudo-Canalejo and coworkers have argued that diffusiophoresis driven by non-specific enzyme-substrate interactions can play a key mechanistic role in enzyme chemotaxis \cite{agudo-canalejo_phoresis_2018}. However, further work is needed to assess whether the nonspecific interactions between an enzyme and its substrate are in fact capable of driving enzyme chemotaxis at observed rates via a diffusiophoretic mechanism. It is also worth noting that the magnitude and direction of nonspecific interactions may depend significantly on whether or not the enzyme has a bound substrate, particularly if the substrate has a nonzero net electrical charge.

\subsection{Mechanisms based on a position-dependent diffusion coefficient}\label{mechanisms position dependent diffusion}
If an enzyme has a higher diffusion coefficient in the presence of its substrate, the diffusion coefficient of the enzyme will be position-dependent in the presence of a substrate concentration gradient. The consequences of a position-dependent diffusion coefficient for experimental observations of chemotaxis are complex and case-dependent (Section \ref{theoretical framework}). For one thing, the transient effect of a position-dependent diffusion coefficient may be different from its steady-state effect. Transiently, enhanced diffusion in the presence of substrate may cause enzyme molecules introduced into the center of a microfluidic channel to diffuse preferentially to a side of the channel that has substrate, relative to a side that contains only buffer (\textbf{Figure \ref{fig1}}), as previously noted \cite{jee_catalytic_2018}. This initial “lurch” in the direction of increasing diffusion coefficient has been termed pseudochemotaxis, because it does not result from a directional preference of the diffusing particle \cite{schnitzer_strategies_1990}, only from the fact that a nonuniform concentration distribution will relax faster where the diffusion coefficient is larger. By contrast, at steady state in the presence of a stable substrate gradient, the distribution of enzyme will depend on the value of $\alpha$. If $\alpha=0$ (isothermal form) then the steady-state concentration profile of the enzyme will be uniform, suggesting no chemotaxis. However, if $\alpha=1$ (Ito form), then the steady-state concentration of the enzyme will be lower where the substrate is at higher concentration (Section \ref{theoretical framework}), matching experimental observations of repulsive chemotaxis \cite{jee_enzyme_2018,jee_catalytic_2018}. In principle, it would also yield attractive chemotaxis, at steady state, if the enzyme diffused more {\em slowly} in the presence of substrate, but this scenario has not been reported. Two microscopic mechanisms for a position-dependent diffusion coefficient and an Ito form FP equation have been put forward.  

One involves catalysis-driven self-propulsion of enzymes \cite{jee_enzyme_2018,weistuch_spatiotemporal_2018}; in effect, an enzyme gets extra propulsive “kicks” in the presence of substrate, leading to an increasing diffusion coefficient along a substrate gradient. This mechanism resembles that of a temperature gradient, where a particle gets more kinetic kicks in regions of higher temperature. Given that diffusion in a temperature gradient can lead to an Ito FP equation \cite{landauer_stability_1983}, it is likely that the self-propulsion mechanism does too. However, the power available from enzyme-catalyzed reactions does not appear to be enough to account for the self-propulsion required by this proposed chemotaxis mechanism (Section\ref{thermodynamics self-propulsion}).  

The other mechanism does not involve self-propulsion, but instead a local equilibrium between substrate-bound and free forms of the enzyme having different diffusion coefficients. A novel derivation from Agudo-Canalejo and coworkers shows that this scenario leads to the Ito form FP equation with a position-dependent mean diffusion coefficient for the enzyme \cite{agudo-canalejo_phoresis_2018}. Because this mechanism does not rely on chemical energy to power diffusive motion, it is not subject to the thermodynamic restriction of the self-propulsion mechanism. 

It would be valuable to carry out more detailed and quantitative calculations that would test the ability of these mechanisms to account for the magnitudes of the effects seen experimentally. Specific questions are whether binding of substrate changes the hydrodynamic radius of urease enough to fit experiment and whether the time scales of experiments showing repulsive chemotaxis \cite{jee_catalytic_2018} are long enough for enzyme molecules to drastically redistribute across the microfluidic channel due merely to space-dependent diffusion.

\section{CONCLUSIONS}

Enhanced diffusion and chemotaxis of enzymes have emerged in recent years as novel phenomena with potential implications in biology and biotechnology. They also pose intriguing puzzles, whose resolution could yield new insights into molecular processes and experimental methods.  Elucidation of the underlying mechanisms will require analysis of potentially subtle linkages among non-equilibrium processes spanning a range of scales. Ultimately,  understanding the mechanisms should make it possible to design enzymes or other molecules to maximize these effects and put them to use.  Continued work in this field promises new insights into the intricacies of molecular motions in out-of-equilibrum systems.

\section{FUTURE ISSUES}
\begin{itemize}
\item What characteristics of enzymes – e.g. structural or catalytic -- correlate with EED and/or chemotaxis? 
\item Are there design principles to discover and even utilize? 
\item Are different mechanisms at work in different enzymes?
\item What role, if any, does the catalytic release of chemical energy play in EED and enzyme chemotaxis?
\item Why do DOSY and DLS yield diffusion results so different from those of FCS for aldolase?  
\item Is enzyme chemotaxis purely a transient phenomenon, or can it also be observed when the enzyme concentration profile is at steady state in a stable substrate gradient?
\item What determines whether a chemotactic enzyme will undergo attractive versus repulsive enzyme chemotaxis?
\end{itemize}

\section*{DISCLOSURE STATEMENT}
MKG has an equity interest in, and is a cofounder and scientific advisor of, VeraChem LLC.

\bibliographystyle{ieeetr}
\bibliography{main}

\end{document}